\documentclass[twocolumn]{aastex63}
\usepackage{multirow}
\usepackage{threeparttable}
\shorttitle{Testing the WEP with the BNS merger GW170817 $\cdots$}
\shortauthors{Yao et al.}
\graphicspath{{./}{figures/}}
\begin{document}

\title{Testing the weak equivalence principle with the binary neutron star merger GW170817: the gravitational contribution of the host galaxy}

\correspondingauthor{Molin Liu}
\email{mlliu@xynu.edu.cn}

\author{Lulu Yao}
\affiliation{College of Physics and Electronic Engineering, Xinyang Normal University, Xinyang 464000, P. R. China}
\author{Zonghua Zhao}
\affiliation{College of Physics and Electronic Engineering, Xinyang Normal University, Xinyang 464000, P. R. China}
\affiliation{Department of Astronomy, Xiamen University, Xiamen, Fujian 361005, P. R. China}
\author{Yu Han}
\affiliation{College of Physics and Electronic Engineering, Xinyang Normal University, Xinyang 464000, P. R. China}
\author{Jingbo Wang}
\affiliation{College of Physics and Electronic Engineering, Xinyang Normal University, Xinyang 464000, P. R. China}
\author{Tong Liu}
\affiliation{Department of Astronomy, Xiamen University, Xiamen, Fujian 361005, P. R. China}
\author{Molin Liu}
\affiliation{College of Physics and Electronic Engineering, Xinyang Normal University, Xinyang 464000, P. R. China}

\begin{abstract}
The successful detection of the binary neutron star (BNS) merger GW170817 and its electromagnetic (EM) counterparts has provided an opportunity to explore the joint effect of the host galaxy and the Milky Way (MW) on the weak equivalence principle (WEP) test. In this paper, using the Navarro$-$Frenk$-$White (NFW) profile and the Herquist profile, we present an analytic model to calculate the galactic potential, in which the possible locations of the source by the observed angle offset and the second supernova (SN2) kick are accounted for. We show that the upper limit of $\Delta \gamma$ is $10^{-9}$ for the comparison between GW170817 and a gamma-ray burst (GRB 170817A), and it is $10^{-4}$ for the comparison between GW170817 and a bright optical transient (SSS17a, now with the IAU identification of AT 2017gfo). These limits are more stringent by one to two orders of magnitude than those determined solely by the measured MW potential in the literature. We demonstrate that the WEP test is strengthened by contribution from the host galaxy to the Shapiro time delay. Meanwhile, we also find that large natal kicks produce a maximum deviation of about $20\%$ to the results with a typical kick velocity 400$\sim$ 500 km s$^{-1}$. Finally, we analyze the impact from the halo mass of NGC 4993 with a typical 0.2 dex uncertainty, and find that the upper limit of $\Delta \gamma$, with a maximum mass $10^{12.4}h^{-1} M_{\odot}$, is nearly two times more stringent than that of the minimum mass $10^{12.0}h^{-1} M_{\odot}$.
\end{abstract}
\keywords{Gravitational waves -- Black hole physics -- Gamma ray bursts -- Binary stars}

\section{Introduction}\label{section1}
On 2015 September 14, the Advanced LIGO detectors picked up the first binary black hole coalescence, GW150914, beginning a new era of observational gravitational-wave (GW) astronomy \citep[][]{GW150914}. Meanwhile, it is believed that the coalescence of a binary neutron star (BNS) system is expected to produce, in addition to GWs, multiple electromagnetic (EM) signatures in different timescales \citep[][]{ENakar,BDMetzger}. For a long time, people had been looking for the EM partners of GWs, but well-accepted result had not been obtained other than a few possible events such as the Fermi Gamma-ray Burst Monitor (GBM) transient 150914 \citep[][]{connaughton}. Then, the big breakthrough came with the detection of GW signal GW170817, which was recorded by the LIGO/Virgo (LIV) GW observatory network on 2017 August 17, 12:41:04 UTC. Later analysis showed that GW170817 was consistent with a BNS inspiral and merger by \citet{LIGO-PRL}. The GW170817 skymap was then released by LIGO/Virgo, thus driving an intensive multi-messenger campaign covering the entire EM spectrum to search for the counterparts of the event \citep[][]{Multi-messenger}. Independently, a gamma-ray signal, classified as a short gamma-ray burst (sGRB), GRB 170817A, coincident in time and sky location with GW170817, was detected using the GBM by \citet{Goldstein} and the International Gamma-Ray Astrophysics Laboratory (INTEGRAL) by \citet{Savchenko}. Beyond the sGRB, multiple independent surveys across the EM spectrum were launched in search of a counterpart. An optical counterpart (OT), Swope Supernova Survey 2017a (SSS17a/AT 2017gfo), was first discovered by the One-Meter Two Hemisphere (1M2H) team in the optical less than 11 hours after merger, associated with NGC 4993 by \citet{Coulter}, a nearby early-type E/S0 galaxy. Five other teams, DLT40 \citep[][]{DLT40}, VISTA \citep[][]{VISTA}, MASTER \citep[][]{MASTER}, DECam \citep[][]{DECam}, and Las Cumbres \citep[][]{LasCumbres}, made independent detections of the same optical transient and host galaxy all within about one hour and reported their results to one another within about five hours. Meanwhile, the source was reported to be offset from the center of NGC 4993 by a projected distance of about $10''$ \citep[][]{Progenitor,Coulter,Levan,Haggard,Kasliwal}, and the binary was determined to potentially lie in front of the bulk of the host galaxy due to the absence of interstellar medium (ISM) absorption in the counterpart spectrum by Hubble Space Telescope (HST) and Chandra imaging, combined with Very Large Telescope (VLT)/MUSE integral field spectroscopy \citep[][]{Levan}. It should be mentioned that the statement above on the discovery of the EM counterpart of GW170817 is not sufficiently convincing, and we direct the reader to the relevant reviews concerning the complete counterpart research of GW170817 \citep[e.g.,][]{Multi-messenger}.

Testing fundamental physics through high-energy astronomical events (HEAE) has always been the subject of research \citep{Will1,Will2}. One famous scheme consists of testing WEP by the comparison of difference waves in HEAE. The pioneering test was that between photons and neutrinos in supernova SN1987A in the Large Magellanic Cloud by \citet{Longo} and \citet{Krauss}. Recently, such schemes have sprung up in physics and astronomy, mainly focusing on cosmic transients such as GRBs \citep[e.g.,][]{Gao}, FRBs \citep[e.g.,][]{JJWei1,SJTingay}, blazar flares \citep[e.g.,][]{JJWei2,ZYWang} and GW event of GW150914 \citep[e.g.,][]{XFWu,Kahya,Liu1}.
\begin{table}\scriptsize
\caption{Localizations in equatorial coordinate system.}
\label{positions}
\centering
\begin{tabular}{llll} 
\hline\hline
\text{Object}&\text{R.A.}&\text{Decl.}&\text{References}\\
\hline
\text{NGC 4993} &13:09:47.7 &-23:23:01&\citet{Coulter}\\
\text{AT 2017gfo}&13:09:48.085 &-23:22:53.343&\citet{Coulter}\\
\text{MW}&17:45:40.04&-29:00:28.1&\citet{Gillessen}\\
\hline\hline
\end{tabular}
\end{table}

After the BNS merger GW170817 and its multiple EM signatures were observed by various astronomical observatories, several pioneering works have presented the WEP tests and produced constraints on the parameterized post-Newtonian (PPN) parameters \citep[][]{LVFI,HWang,JJWei3}. \citet{LVFI} constrained the deviation of the speed of gravity, and violations of Lorentz invariance and the equivalence principle are presented by the observed temporal offset, the distance to the source, and the assumed emission time difference, in which the bound on the difference of $\gamma_{\text{GW}} - \gamma_{\text{EM}}$ was given in the range of $[-2.6\times10^{-7}, 1.2\times10^{-6}]$. Then by assuming the simultaneous emission of GWs and photons, \citet{HWang} presented a result of $\Delta \gamma \leq 10^{-7}$, which could be improved to $4 \times 10^{-9}$ using the potential fluctuations from large-scale structure, as originally proposed by \citet{Nusser}. Meanwhile, \citet{JJWei3} considered a Keplerian potential $\Phi = -G M/r$ for two cases: the MW and the Virgo Cluster. The former adopted a total mass of $6\times 10^{11} M_{\odot}$ and gave the upper limits $\sim 10^{-8}$ for GW170817/GRB 170817A and $\sim 10^{-3}$ for GW170817/AT 2017gfo.

Meanwhile we have noticed that according to the K-band luminosity in the 2MASS Redshift Survey \citep[see][]{Huchra}, the stellar mass of NGC 4993 ($\sim 6.2\times10^{10} M_{\odot}$) is almost equal to that of the MW ($\sim 6.4\times10^{10} M_{\odot}$). Therefore, it can be expected that the gravitational effect from the host galaxy will largely enhance the WEP test when comparing with the tests that only consider the MW. As far as we known, no works refer to the test involving NGC 4993. Motivated by this situation, we restudy the WEP test of GW170817, but consider a joint gravitational potential that consists of the host galaxy and the MW. In this work, we focus on three aspects of the tests: the observed angle offset from the source, the possible large natal kick on the BNS, and the typical uncertainty, 0.2 dex, on the halo mass of NGC 4993.

The outline of this paper is as followings. In Section \ref{section2}, we present a computable galactic model by considering the observed angle offset. In Section \ref{section3}, we obtain the constraints on the WEP test via the joint potential. We then explore the impacts of the large natal kick and the halo mass of NGC 4993 on the tests in Sections \ref{section4} and \ref{section5}, respectively. Section \ref{section6} presents the conclusion.
\section{Model including the joint gravitational effect of the MW and NGC 4993}\label{section2}
\subsection{Traveling path of waves from the merge position}\label{section2a}
The gravitational potential driving waves traveling in interstellar space can be divided into three parts: the MW $\Phi_{mw}$, a flat intergalactic background $\Phi_{ig}$, the host galaxy $\Phi_{host}$ \citep[see][]{Gao}. In previous WEP tests, the latter two potentials are usually neglected due to the comparative lack of observations of the source. However, GW170817 provides us with some important observations about the host galaxy. We thus follow the observations of the source and try to build a computable galactic model to calculate the Shapiro time of the traveling waves. The geometry of our model about the traveling path of the waves from the merge position are given in Appendix \ref{app-1}. Localizations of the centers of the MW, NGC 4993, and the source are listed in Table \ref{positions} by using J2000.0 \citep[see][]{Gillessen}, where the equatorial coordinate system (ECS) coordinates are used. Meanwhile, considering the observations of the ISM absorption \citep[see][]{Levan}, the source is located in the region bounded by the observational angle offset (see the shaded area in Figure \ref{fig1-2}).
\begin{figure*}
\centering
\includegraphics[width=\columnwidth]{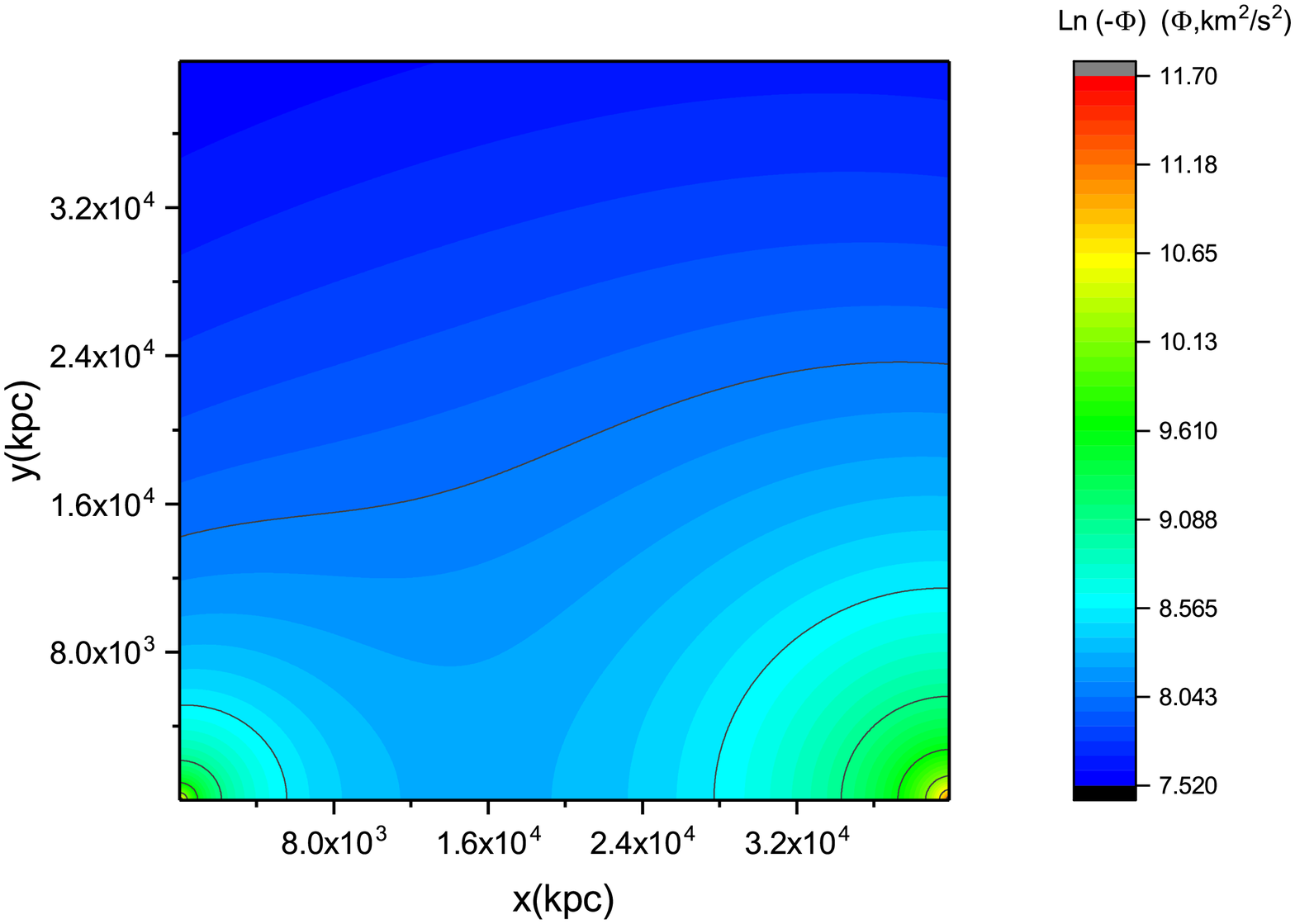}
\includegraphics[width=0.8\columnwidth]{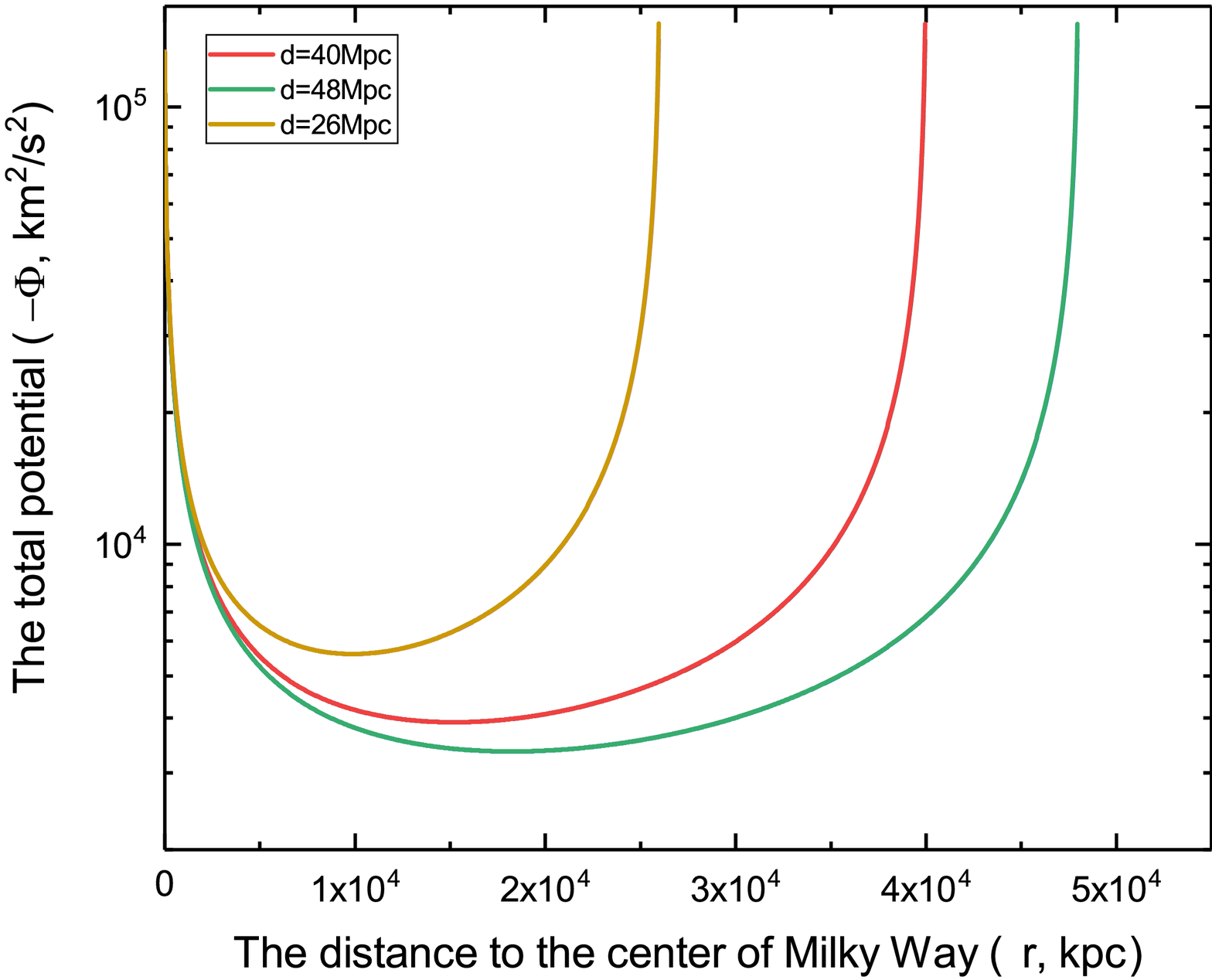}
\caption{\scriptsize The joint gravitational potential consisting of the MW and the NGC 4993. The left panel shows the contour plots of the potential. The right panel shows the potential along the waves path. The median magnitude of luminosity distance is $d = 40$~Mpc}
\label{fig3:potential}
\end{figure*}
\begin{table}\scriptsize
\caption{NFW DM halo parameters.}
\label{NFWparameters}
\centering
\begin{tabular}{lll} 
\hline\hline
NFW Parameters&MW&NGC 4993\\
\hline
median $r_{200}$~(kpc)&288&282\\
concentration parameter $c_{200}$&5.8&5.9\\
density parameter $\rho_{0}$ ($10^{-3} M_{\odot} pc^{-3}$)&1.6&1.6\\
scale radius $R_{s}$~(kpc)&49&48\\
\hline\hline
\end{tabular}
\end{table}
\subsection{Gravitational potential by the joint effect of the MW and NGC 4993}\label{section2b}
In this work, the enclosed masses consist of the stellar mass and the dark matter (DM) halo through the spherically symmetric profiles. The former is described by a Hernquist profile \citep[][]{Hernquist}, and the latter is described by a NFW profile \citep[][]{Navarro}. The density distribution of the stellar component was given by \citet{Hernquist} as follows:
\begin{equation}\label{21Hernquist}
   \rho_{s}(r) = \frac{M_{s} a_{b}}{2 \pi r \left(r + a_{b}\right)^3},
\end{equation}
where $M_{s}$ is the total stellar mass and $a_{b}$ is a scale length. The potential is thus given by
\begin{equation}\label{21stelpoten}
\Phi_{\text{stellar}} (r) = - \frac{G M_{s}}{r + a_{b}}.
\end{equation}
The stellar mass of MW is $6.4 \times 10^{10} M_{\odot}$ given by \citet{McMillan}, and the stellar mass of NGC 4993 is $6.2 \times 10^{10} M_{\odot}$ provided by \citet{Lim}. The bulge scale length is $0.5$~kpc for the MW given by \citet{Sofue}. The bulge scale length of NGC 4993 is about $0.55$ times the half-light radius $R_{\text{eff}}$ \citep[][]{Hernquist}, which was observed recently as $15.''5\pm1.''5$, which corresponds to a $3.0$~kpc offset for a distance of 40~Mpc, using HST measurements \citep[][]{Hjorth}.

The density distribution of the DM halo component was given by \citet{Navarro} as
\begin{equation}\label{22NFW}
  \rho_{\text{DM}}(r) = \frac{\rho_0 R_s}{r}\left(1 + \frac{r}{R_s}\right)^{-2},
\end{equation}
where $\rho_0$ is the density parameter, and $R_s$ is the scale radius defined by $R_s = R_{200}/c_{200}$. $R_{200}$ is the position at which the enclosed density is $200$ times the universe's critical density. $c_{200}$ is the concentration parameter obtained via the empirical expression given by \citet{Duffy}
\begin{equation}\label{EEofDuffy}
\log_{10}c_{200} = (0.76\pm _{0.01}^{0.01}) + (-0.10\pm _{0.01}^{0.01})\log_{10} \left(\frac{M_{200}}{M_{\text{pivot}}}\right),
\end{equation}
where the median halo mass $M_{\text{pivot}} = 2\times 10^{12} h^{-1} M_{\odot}$. Based on the report from \citet{Planck}, the median value for the Hubble parameter is $h = 0.679$. The halo mass of the MW is adopted as 2.5 $\pm$ 1.5 $\times$ 10$^{12}$ M$_{\odot}$ obtained from the numerical action method by \citet{Phelps}. The halo mass of NGC 4993 is adopted as $(10^{12.2} h^{-1}) M_{\odot}$ obtained from the 2MASS Redshift Survey (2MRS) in the low redshift universe by \citet{Lim}. Therefore, the parameters ($\rho_{0}$, $R_{s}$, $R_{200}$ and $c_{200}$) can be obtained through modeling the NFW halo, and they are listed in Table \ref{NFWparameters}. The potential of the NFW halo is thus given by
\begin{equation}\label{22NFWpoten}
  \Phi_{\text{DM}} (r) = - \frac{4 \pi G \rho_0 R_s^3}{r}\ln \left(1 + \frac{r}{R_s}\right).
\end{equation}

Based on the main components of the stellar and DM halos, the total potential $\Phi_{\text{total}}$ can be given as,
\begin{equation}\label{addtotalpo1}
\Phi_{\text{total}} = \Phi_{\text{mw}} + \Phi_{\text{host}},
\end{equation}
where the potential $\Phi_{\text{mw}}$ of MW (or $\Phi_{\text{host}}$ of NGC 4993) is composed of the Hernquist stellar sector $\Phi_{\text{s1}}$ (or $\Phi_{\text{s2}}$ of NGC 4993) from Equation (\ref{21stelpoten}) and the NFW halo sector $\Phi_{\text{D1}}$ (or $\Phi_{\text{D2}}$ of NGC 4993) from Equation (\ref{22NFWpoten}). Therefore, $\Phi_{\text{mw}}$ and $\Phi_{\text{host}}$ are shown by
\begin{eqnarray}
\label{addmw}\Phi_{\text{mw}} &=& \Phi_{\text{s1}} (r) + \Phi_{\text{D1}} (r),\\
\label{addNGC}\Phi_{\text{host}} &=& \Phi_{\text{s2}} (\chi) + \Phi_{\text{D2}} (\chi),
\end{eqnarray}
where $\chi(r, \theta)$ is given in formula (\ref{addppnn}) and refers to a dynamical distance from the center of NGC 4993 to the point on the traveling path of the waves. The total potential $\Phi_{\text{total}}$ is illustrated in Figure \ref{fig3:potential}. The left panel shows the profile, and the median magnitude of the luminosity distance $d = 40$~Mpc is adopted. The right panel shows the path considering the condition of Equation (\ref{addppnn}) where the observed luminosity distance of $d = 40^{+8}_{-14}$~Mpc is adopted. The two panels strongly suggest that the impacts of the host galaxy on the total potential should not be ignored. The model parameters for the Hernquist stellar profile both in the MW and NGC 4993 are given by Equation (\ref{21stelpoten}), and the parameters of the NFW halo are listed in Table \ref{NFWparameters}.

The Shapiro time delay $\Delta t_{gra}$ can be obtained through the integration of the potential along the path \citep[e.g.,][]{Shapiro,Longo,Krauss}
\begin{equation}\label{TD-1}
    \Delta t_{\text{gra}} = - \frac{1 + \gamma}{c^3} \int_{r_{e}}^{r_{o}} \Phi(r) d r,
\end{equation}
where $r_e = r_S$ and $r_o = r_G$ denote the positions of sender and receiver. Meanwhile, in order to define the waves traveling along the path from the merge position to the Earth, the condition of Equation(\ref{addPN}) must be upheld. Due to the possible large natal kick of the binary \citep[see][]{Progenitor}, the source is possibly kicked outside of the gravitational grasp of NGC 4993. The angle offset and the large natal kick thus become the major factors affecting the position of the transient in the WEP test.
\section{WEP test of the binary neutron star merger GW170817 with the angle offset}\label{section3}
\subsection{Constraints on the WEP test between GW170817 and GRB 170817A}\label{subsection31}
In our tests, GRB 170817A and AT 2017gfo are substituted into the calculations, acting as the counterparts of GW170817. The temporal offset between the BNS merger and GRB is $1.734 \pm 0.054$~s, which can be treated as the maximum time delay caused by the gravitational potential. Substituting the Shapiro time into Equation (\ref{TD-1}), we obtain the upper limit of the PPN parameter difference between GW170817 and GRB 170817A in the potentials of the two galaxies, denoted by $|\Delta\gamma_1| \equiv |\gamma_{\text{GW}} - \gamma_{\text{EM}}|$. Here, two extreme positions of the transient are considered: one is located at the projected point of the center of NGC 4993 and the other is located near the edge of the galaxy. Appendix \ref{app-2} gives details of the geometry of the angle offset and these extreme positions of transient.

The results are listed in Table \ref{tabloffset} (see $\gamma_1$) and primarily show that the difference $|\Delta \gamma_1|$ between GW170817 and GRB 170817A is under $10^{-9}$ due to the joint effect of the MW and NGC 4993. When comparing with previous results, which only accounted for the MW potential, our result is more stringent by two orders of magnitude than the result of $10^{-7}$, using the method of the impact parameter from \citet{HWang}. Additionally, it is also more stringent by one order of magnitude than the result of $10^{-8}$ obtained via the Keplerian potential method from \citet{JJWei3}. The total mass of the MW was adopted as $6 \times 10^{11} M_{\odot}$ in both methods. When the gravitational contribution of NGC 4993 was added to the WEP test, the mass of the galaxies ($\sim 10^{12} M_{\odot}$) is larger than that of the MW ($\sim 10^{11} M_{\odot}$) adopted before. Therefore, our results are enhanced significantly by one to two order of magnitudes compared to those of only considering the MW when adding the contribution of the host galaxy to the tests.

Another advantage of exploring the host galaxy is that it provides us with an alternative to alleviate the suppression of the WEP constraint caused by the integral of the potential far beyond the MW. The contribution to the test caused by the alteration of source location is highly suppressed for the MW (see lines 1 and 4 in Table \ref{tabloffset}). However, this kind of suppression can be alleviated when the test contains the host galaxy. If the location of the transient changes from the edge of the galaxy to the center, the deviation $\delta_1$ is positive, and the constraints are more stringent by about $1\%$ (see lines 2, 3, 5, and 6 in Table \ref{tabloffset}).
\begin{table*}
\caption{Upper limits of the PPN parameter differences for three kinds of enclosed mass.} \label{tabloffset}
\begin{threeparttable}
\begin{tabular*}{\hsize}{@{}@{\extracolsep{\fill}}cllll@{}}
\hline\hline
Comparison Type&$r_S = GN'$\tnote{c}&$r_S = GR_N$\tnote{d}&$\delta_1$\tnote{e}&Enclosed mass\\
\hline
\multicolumn{1}{l}{\multirow{3}{*}{GW170817/GRB170817A ($|\Delta\gamma_1|\lesssim$)}}&$6.1_{-0.3}^{+0.8} \times 10^{-9}$&$6.1_{-0.3}^{+0.8} \times 10^{-9}$&0.0\%&MW\tnote{b}\\
&$6.5_{-0.3}^{+0.8} \times 10^{-9}$&$6.5_{-0.3}^{+0.9} \times 10^{-9}$&1.3\%&NGC 4993\\
&$3.1_{-0.2}^{+0.4} \times 10^{-9}$&$3.2_{-0.2}^{+0.4} \times 10^{-9}$&0.6\%&MW + NGC 4993\tnote{a}\\
\hline
\multicolumn{1}{l}{\multirow{3}{*}{GW170817/AT2017gfo ($|\Delta \gamma_2|\lesssim$)}}&$1.4_{-0.1}^{+0.2} \times 10^{-4}$&$1.4_{-0.1}^{+0.2} \times 10^{-4}$&0.0\%&MW\tnote{b}\\
&$1.5_{-0.1}^{+0.2} \times 10^{-4}$ &$1.5_{-0.1}^{+0.2} \times 10^{-4}$&1.3\%&NGC 4993\\
&$7.1_{-0.3}^{+0.9} \times 10^{-5}$&$7.1_{-0.4}^{+0.9} \times 10^{-5}$&0.6\%&MW + NGC 4993\tnote{a}\\
\hline\hline
\end{tabular*}
\begin{tablenotes}
\footnotesize
\item[] \textbf{Note.} The constraints of the WEP tests are calculated through two kinds possible source locations $r_S$ = $GN'$(or $GR_N$) by taking into account the observed angle offset \citep[][]{Progenitor,Coulter} and the absence of ISM absorption in the counterpart spectrum \citep[][]{Levan}.
\item[a] The test of the maximum enclosed mass is calculated via the total potential $\Phi_{\text{total}}$ (\ref{addtotalpo1}).
\item[b] The test of only the MW is calculated via the potential $\Phi_{\text{MW}}$ (\ref{addmw}), and the disappeared deviation $\delta_1$ indicates that the impact of the source location in NGC 4993 on the test is almost entirely suppressed.
\item[c] It corresponds the maximum propagation distance, and the source is located at the projected point $N'$ from the center of NGC 4993 (see Figure \ref{fig1-2}).
\item[d] It corresponds to the minimum propagation distance, and the source is located at the point $R_N$ near the edge of NGC 4993.
\item[e] The influence of the change of source position on the WEP test is quantified through the deviation $\delta_1$ defined by $\delta_1 = \left[\Delta \gamma (G R_N)-\Delta \gamma (G N')\right]/\Delta \gamma (G N')$. A positive value of $\delta_1$ means that the constraint on the PPN parameter becomes tighter when the source position changes from the edge to the center of NGC 4993.
\end{tablenotes}
 \end{threeparttable}
\end{table*}
\subsection{Constraints on the WEP test between GW170817 and AT 2017gfo}
The observations of optical source show that the time difference between GW170817 and AT 2017gfo is $10.87$~hr. If we treat the time offset as the Shapiro time delay, the upper limits of $\Delta \gamma_2$ in the comparison of GW170817/AT 2017gfo are obtained by considering two kinds positions of source. The test results are listed in Table \ref{tabloffset} (see $\gamma_2$).
\begin{table*}
\caption{Upper limits of $\Delta \gamma$ from the large natal kick.} \label{reladev}
\begin{threeparttable}
\begin{tabular*}{\hsize}{@{}@{\extracolsep{\fill}}clllll@{}}
\hline\hline
d&$V_{\text{kick}}$&$\tau_{\text{gwr}}$&$R_{\text{real}}$&Upper limit of $\Delta\gamma_1$&Upper limit of $\Delta\gamma_2$\\
\hline
\multicolumn{1}{l}{\multirow{4}{*}{26~Mpc}}&400~km/s&$t_{\text{Hubble}}$&2.7~Mpc&$3.7\times10^{-9}$&$8.3\times10^{-5}$\\
&500~km/s&$t_{\text{Hubble}}$&3.4 Mpc&$3.7\times10^{-9}$&$8.3\times10^{-5}$\\
&400~km/s&86~Myr&36~kpc&$3.5\times10^{-9}$&$8.0\times10^{-5}$\\
&500~km/s&86~Myr&45~kpc&$3.5\times10^{-9}$&$8.0\times10^{-5}$\\
\hline
\multicolumn{1}{l}{\multirow{4}{*}{48~Mpc}}&400~km/s&$t_{\text{Hubble}}$&2.7~Mpc&$3.0\times10^{-9}$&$6.8\times10^{-5}$\\
&500~km/s&$t_{\text{Hubble}}$&3.4~Mpc&$3.0\times10^{-9}$&$6.9\times10^{-5}$\\
&400~km/s&86~Myr&36~kpc&$3.0\times10^{-9}$&$6.7\times10^{-5}$\\
&500~km/s&86~Myr&45~kpc&$3.0\times10^{-9}$&$6.7\times10^{-5}$\\
\hline\hline
\end{tabular*}
\begin{tablenotes}
\footnotesize
\item[] \textbf{Note.} The WEP constraints are calculated by the total potential (\ref{addtotalpo1}) with the maximum enclosed mass scale (MW + NGC 4993). The merger time of the BNS and the kick velocity are taken from \citet{Tauris} and \citet{Progenitor}.
\end{tablenotes}
\end{threeparttable}
\end{table*}

The results show that the differences $\Delta \gamma_2$ are all under an order of magnitude of $10^{-4}$ in three scales, and the WEP test of GW170817/AT 2017gfo is significantly enhanced by the host galaxy. Even for the enclosed mass of the MW, the result of $3.4 \times 10^{-4}$ is more stringent by one order of magnitude than the limit of $1.4 \times 10^{-3}$ in the Keplerian potential given by \citet{JJWei3}. When comparing with that of GW170817/GRB 170817A in Section \ref{subsection31}, the result of GW170817/AT 2017gfo is less stringent by four or five orders of magnitude, which means that the bound on the observed delay is weaker for the comparison of GW170817/AT 2017gfo.
\section{WEP test of the binary neutron star merger GW170817 with the large natal kick}\label{section4}
The actual distance to the final merger is also strongly influenced by the SN2 kick. According to the kinematic modeling from SN2 to the merger, the slingshot effect caused by the tangential SN2 kick is much more efficient than a purely radial kick driving the binary to the outer regions of the galaxy \citep[see][]{Progenitor}. Therefore, a large natal kick to the binary could make it merge at a greater distance. The final merger position is possibly out of range of the galaxy for a larger SN2 kick, as long as the observed offset angle is respected. Therefore, the merger position $S$ may be out of range of $|N'R_{N}|$ (see Figure \ref{fig1-2}). The real distance $R_{\text{real}}$ from the SN2 to the merger can be simplified as follows:
\begin{equation}\label{Rreal}
R_{\text{real}} = \tau_{\text{gwr}} V_{\text{kick}},
\end{equation}
where $\tau_{gwr}$ is the merger time of the BNS, and $V_{\text{kick}}$ is the kick velocity along the radial direction. The merger time of the BNS $\tau_{gwr}$ is in the range of
\begin{equation}\label{admertime}
\tau_{gwr0} \lesssim \tau_{gwr} \lesssim t_{\text{Hubble}}
\end{equation}
where $t_{\text{Hubble}} = 1/H_0$ with $H_0$ = 100 h~km~s$^{-1}$ Mpc$^{-1}$ is the Hubble time, and $\tau_{gwr0}$ is the minimum merger time 86~Myr from the observation of PSR J0737-3039A/B in a highly relativistic orbit \citep[see][]{Tauris}. Meanwhile, the kick velocity $V_{\text{kick}}$ is assumed to be constant after the SN2, whereas there is ample observational evidence for large NS kicks (typically 400$\sim$500 km~s$^{-1}$) in observations of young radio pulsars. Therefore, the distance of the binary after the SN2 can be estimated as $(36-45)\text{kpc}\lesssim R_{\text{real}}\lesssim(2.7-3.4)\text{Mpc}$. It is apparently beyond the diameter of the galaxy NGC 4993, 26~kpc \citep[see][]{Lauberts}. The constraints from the large natal kick on the WEP test are thus obtained and listed in Table \ref{reladev}, in which the case of the maximum enclosed mass, i.e., the scale of MW + NGC 4993, is considered and the perturbation of distance comes from the kick distance $R_{\text{real}}$. Because the transient location is most possibly directly in front of NGC 4993, the travel path will be reduced after a larger natal kick, compared to the calculation without kick. The upper limit of $\Delta \gamma$ is thus less stringent by about 2$\%-$4$\%$ than the cases without the kick, in Section \ref{section3}.

The results show that the maximum upper limit of $\Delta \gamma$ comes from the case where the traveling waves have a maximum kick speed $V_{\text{kick}} \sim$ 500~km~s$^{-1}$ within the Hubble time, and the traveling distance is reduced by 3.4~Mpc. In this case, the upper limit $3.7 \times 10^{-9}$ thus increases by nearly $20\%$ compared to the result of $3.2 \times 10^{-9}$ without the kick effect in Table \ref{tabloffset}. This shows that the large natal kick has a significant impact on the WEP tests.
\section{Influence of the halo mass of NGC 4993 with a 0.2 dex scatter on the WEP test}\label{section5}
\begin{table}\scriptsize
\centering
\caption{NFW halo parameters in NGC 4993.}
\label{addNFWp}
\centering
\begin{tabular}{lcc} 
\hline\hline
NFW Parameters  &Upper limit& Lower limit\\
\hline
halo mass $M_{\text{DM}}$ ($M_{\odot}/h$)&$10^{12.4}$&$10^{12.0}$\\
median $r_{200}$ (kpc) &328&243\\
concentration parameter $c_{200}$&5.6&6.2\\
density parameter $\rho_{0}$ ($10^{-3} M_{\odot} pc^{-3}$)&1.5&1.8\\
scale radius $R_{s}$ (kpc)&58&39\\
\hline
\hline
\end{tabular}
\end{table}

\begin{figure*}
\centering
\includegraphics[width=0.68\columnwidth]{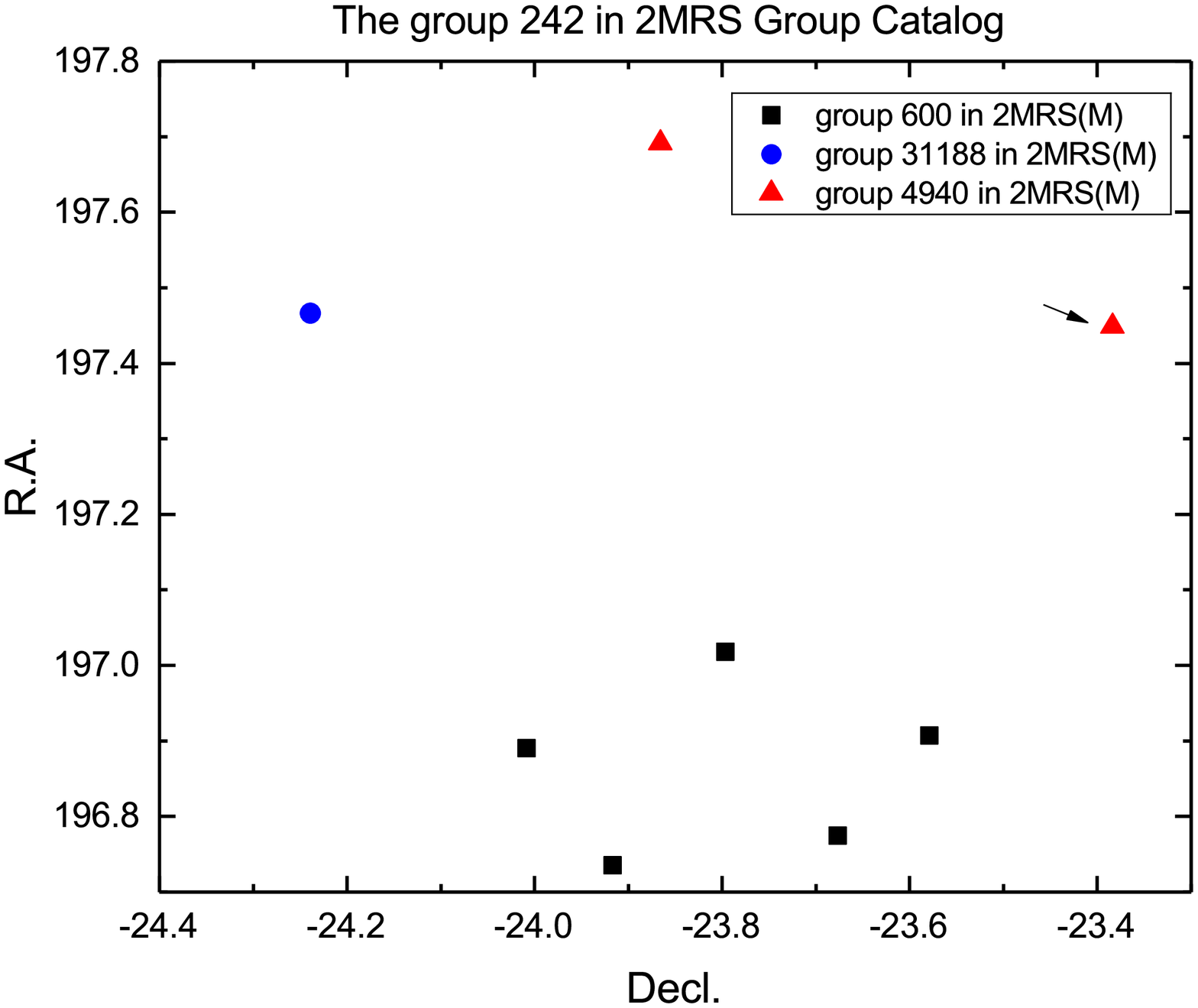}
\includegraphics[width=0.68\columnwidth]{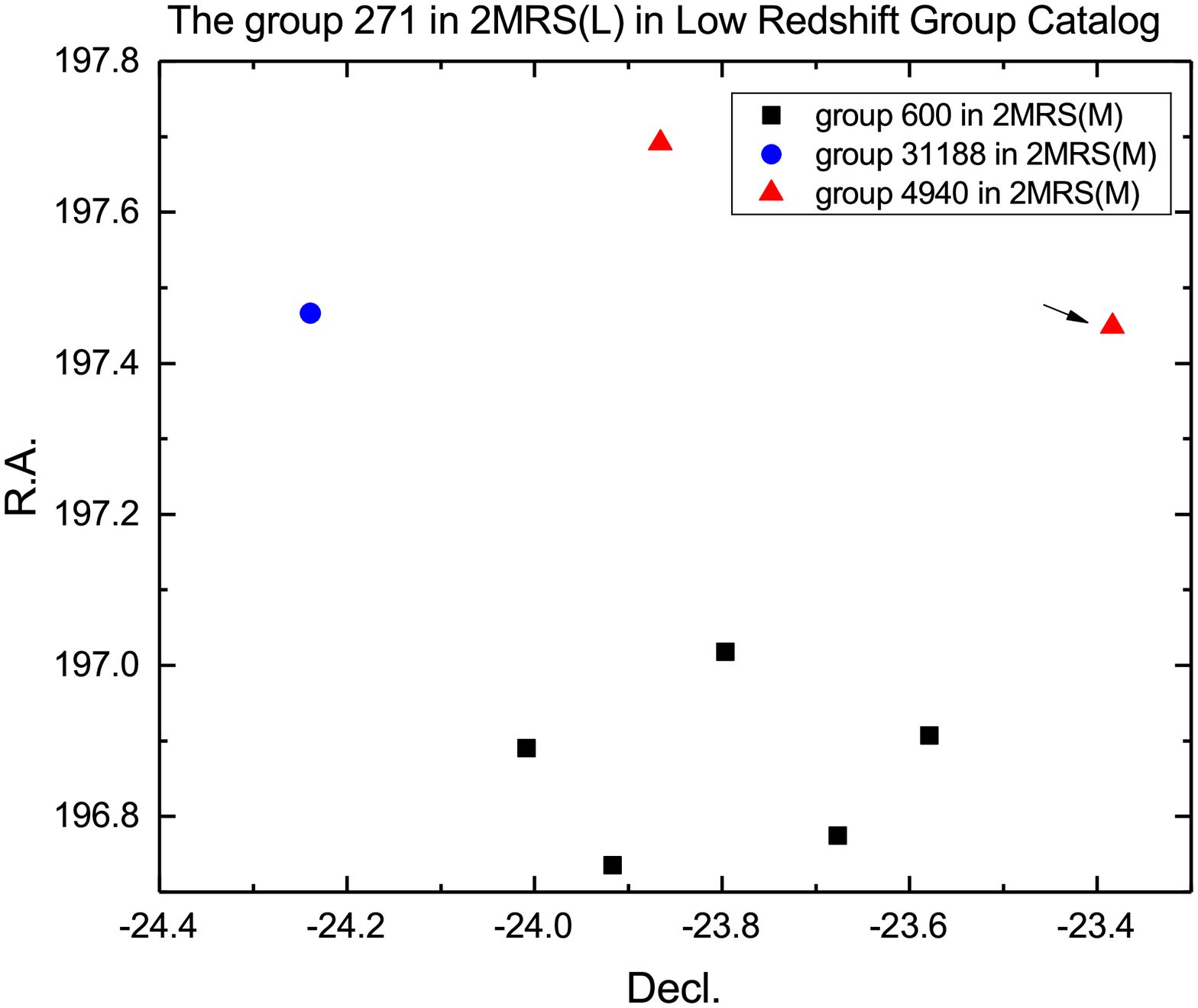}\\
\includegraphics[width=0.68\columnwidth]{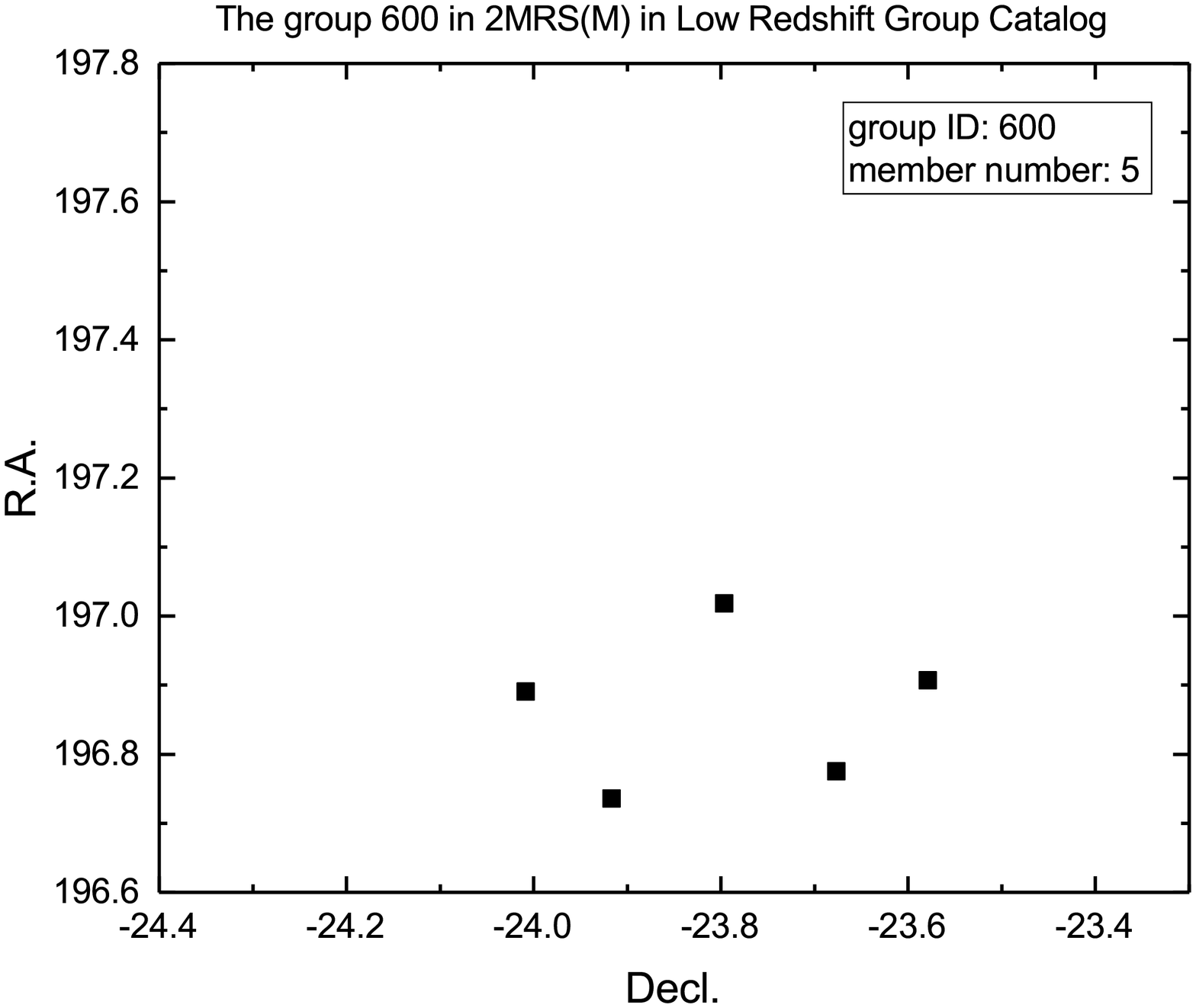}
\includegraphics[width=0.68\columnwidth]{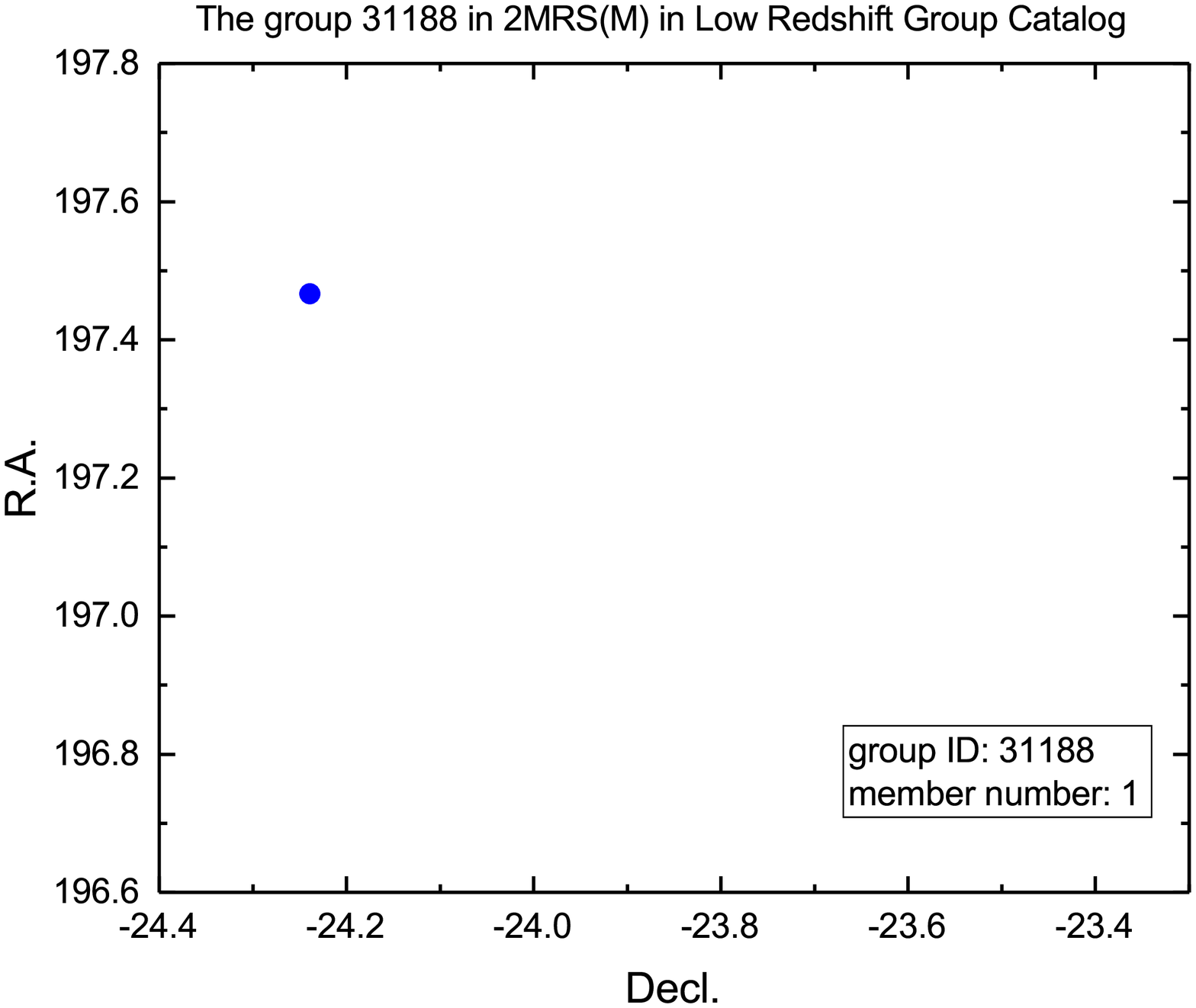}
\includegraphics[width=0.68\columnwidth]{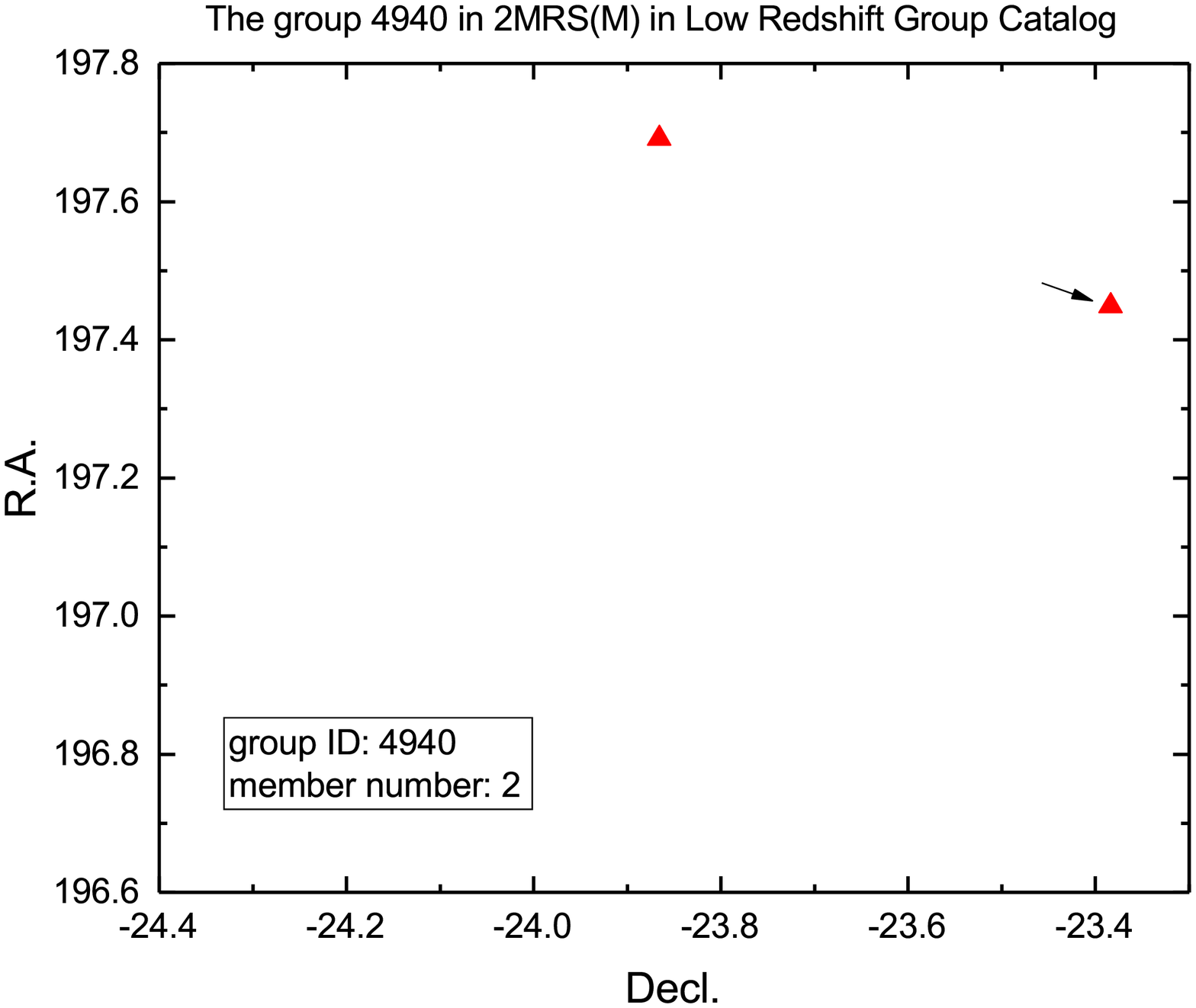}
\caption{\scriptsize The NGC 4993 (indicated by the arrows) in various group catalogs. The group comprised of eight galaxies did not change in the 2MRS Group Catalog (top left, group 242 of 2MRS) nor in 2MRS(L) in the Low Redshift Group Catalog (top right, group 271 of 2MRS(L)). However, group 271 of 2MRS(L) was split into three small groups in the 2MRS(M) in the Low Redshift Group Catalog: the group 600 (bottom left), the group 31188 (bottom middle), and the group 4940 (bottom right). With the group catalog reinforced and the number members decreased, the group thus contains more reliable information about the galaxy, particularly for the groups consisting of one member or a small number of members.}
\label{figDM:group}
\end{figure*}

In Lim's catalogs \citep[see.][]{Lim}, the 2MRS(M) of the Low Redshift Group Catalog was given by using the Proxy-M to estimate halo masses of the galaxies. In 2MRS(M), one can find that the group of NGC 4993 is a poor system that consists of only two galaxies. The group is inside the region of completeness for a given halo mass, and thus we can assign halo mass by abundance matching. In Figure \ref{figDM:group}, we present five panels to illustrate how the related groups change in three kinds of catalogs: the 2MRS Group Catalog provided by \citet{Luy}, the 2MRS(L), and the 2MRS(M) in the Low Redshift Group Catalog. We find that the group will become a poor system, with a decreasing number of members, and the properties of the group would be close to those of the galaxies. Therefore, it is reasonable to use the halo mass $10^{12.2}/h~M_{\odot}$ in the poor system to identify that of galaxy NGC 4993.

Because the halo masses assigned by the group finder are unbiased with respect to the true halo masses, but have a typical uncertainty of $\sim 0.2$~dex in the catalog of 2MRS(M). The halo mass will change, and its range then becomes $[10^{12.0} h^{-1} M_{\odot}, 10^{12.4} h^{-1} M_{\odot}]$. The related NFW halo parameters are listed in Table \ref{addNFWp}. By using these new parameters, the upper limits of $\Delta \gamma_1$ and $\Delta \gamma_2$ are recalculated in Table \ref{tablscatter}. It is thus clear that the tests of the maximum halo mass are nearly two times more stringent than those of the minimum halo mass. This means that the influence of the halo mass of NGC 4993 on the results of the WEP test is significant.
\begin{table*}
\caption{Upper limits of $\Delta \gamma$ with the maximum and minimum halo masses in NGC 4993.} \label{tablscatter}
\begin{threeparttable}
\begin{tabular*}{\hsize}{@{}@{\extracolsep{\fill}}clllll@{}}
\hline\hline
Comparison Type&Maximum halo mass&Minimum halo mass&$\delta_2$&Enclosed mass\\
\hline
\multicolumn{1}{l}{\multirow{2}{*}{GW170817/GRB170817A ($\Delta \gamma_1$)}}&$4.2_{-0.2}^{+0.6} \times 10^{-9}$&$9.8_{-0.5}^{+1.2} \times 10^{-9}$&133.3\%&NGC 4993\\
&$2.5_{-0.1}^{+0.3} \times 10^{-9}$&$3.8_{-0.2}^{+0.5} \times 10^{-9}$&52.1\%&MW + NGC 4993\\
\hline
\multicolumn{1}{l}{\multirow{2}{*}{GW170817/AT2017gfo ($\Delta \gamma_2$)}}&$9.5_{-0.5}^{+1.3} \times 10^{-5}$ &$2.2_{-0.1}^{+0.3} \times 10^{-4}$&131.6\%&NGC 4993\\
&$5.6_{-0.3}^{+0.8} \times 10^{-5}$&$8.5_{-0.4}^{+1.1} \times 10^{-5}$&51.8\%&MW + NGC 4993\\
\hline
\hline
\end{tabular*}
\begin{tablenotes}
\footnotesize
\item[] \textbf{Note.} The halo mass is in the range of $[10^{12.0}h^{-1} M_{\odot}, 10^{12.4}h^{-1} M_{\odot}]$ due to a typical uncertainty of 0.2 dex in NGC 4993. The influence of the change of halo mass on the WEP test is quantified through the deviation $\delta_2$ defined by $\delta_2 = \left[\Delta \gamma (Min)-\Delta \gamma (Max)\right]/\Delta \gamma (Max)$. A positive value of $\delta_2$ means that the constraint on the PPN becomes tighter when the halo mass increases.
\end{tablenotes}
\end{threeparttable}
\end{table*}

\section{Conclusion}\label{section6}
In this paper a model was developed to describe the augmented test of a host galaxy that considers angle offset, large natal kicks, and the typical uncertainty on the halo mass of NGC 4993.

The transient could be located at any point along the line of sight in the NGC 4993 due to the angle offset from the center, as long as the observed luminosity distance is guaranteed. Because the transient is most likely directly in front of the bulk of the host galaxy, the minimal distance of the source from the center of its host is then simply the projected distance, and the maximal distance extends near the edge of NGC 4993. The influence of the angle offset on the results can be quantified by the distance offset shown by $SN'$ in Figure \ref{fig1-2}. The results of the maximum and the minimum $SN'$ are listed in Table \ref{tabloffset}.

The luminosity distance adopted in the calculation is $40_{-14}^{+8}$~Mpc, which was the closest observed GW source and the closest short $\gamma$-ray burst, with a distance measurement by \citet{LIGO-PRL}. Meanwhile, several other EM methods have given more precise values for the distance, e.g., $40.4\pm3.4$~Mpc using the MUSE/VLT measurement of the heliocentric redshift, $41.0\pm3.1$~Mpc using HST measurements of the effective radius and the MUSE/VLT measurements of the velocity dispersion \citep[][]{Hjorth}, and $40.7\pm1.4\pm1.9$~Mpc using surface brightness fluctuations \citep[][]{Cantiello}. Although the uncertainty of $40_{-14}^{+8}$~Mpc is slightly worse than these recent values shown above, it is accurate enough for the WEP test due to the suppression of the host galaxy.

The natal kick imparted to the binary at the same time of the SN explosions that gave rise to the neutron stars. This kind of kick should then lead to mergers at large offsets from their birth sites and host galaxy, on scales of about tens to hundreds of kiloparsecs, over a broad range of merger timescales \citep[][]{Berger}. In this work, we chose large NS kicks with $V_{\text{kick}}$ typically from 400$\sim$500~km~s$^{-1}$. The related delay time was adopted in a less stringent range from the observed minimal magnitude 86~Myr to the Hubble time. For the more stringent constraints on the delay time, one can refer to Figure 8 and Table 2 given in \citet{Progenitor}, where the summary statistics for output PDFs and the more detailed PDFs on progenitor properties, with various delay time constraints, are presented.

\section*{Acknowledgments}
We thank Dr. Jielei Zhang for helpful discussions. This work is supported by the National Natural Science Foundation of China under grant Nos. 11475143 and 11822304, and the Nanhu Scholars Program for Young Scholars of Xinyang Normal University.

\appendix
\section{Traveling path of the waves from NGC 4993 to the MW}\label{app-1}
\begin{figure}
\centering
\includegraphics[width=0.35\columnwidth]{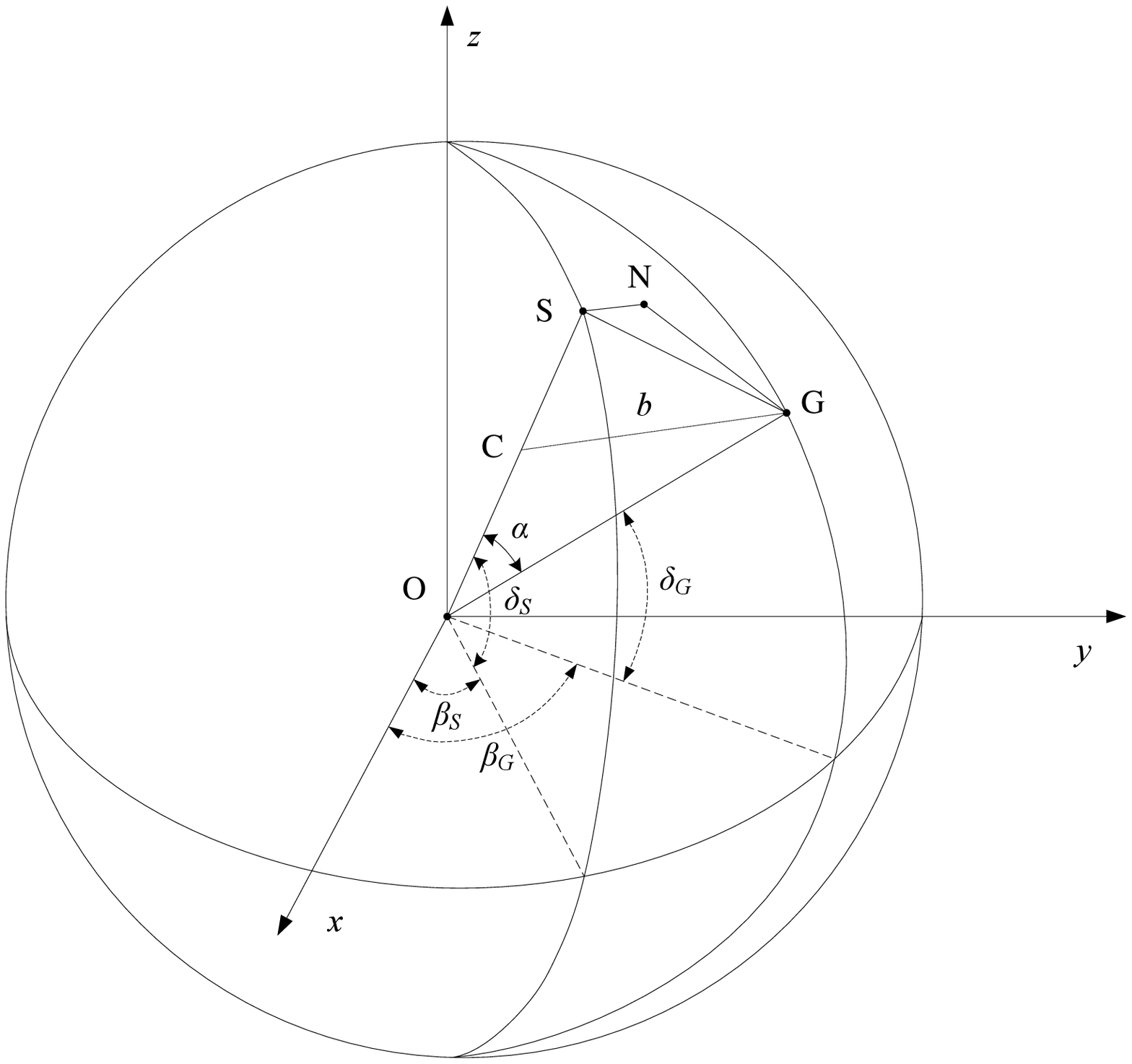}
\includegraphics[width=0.35\columnwidth]{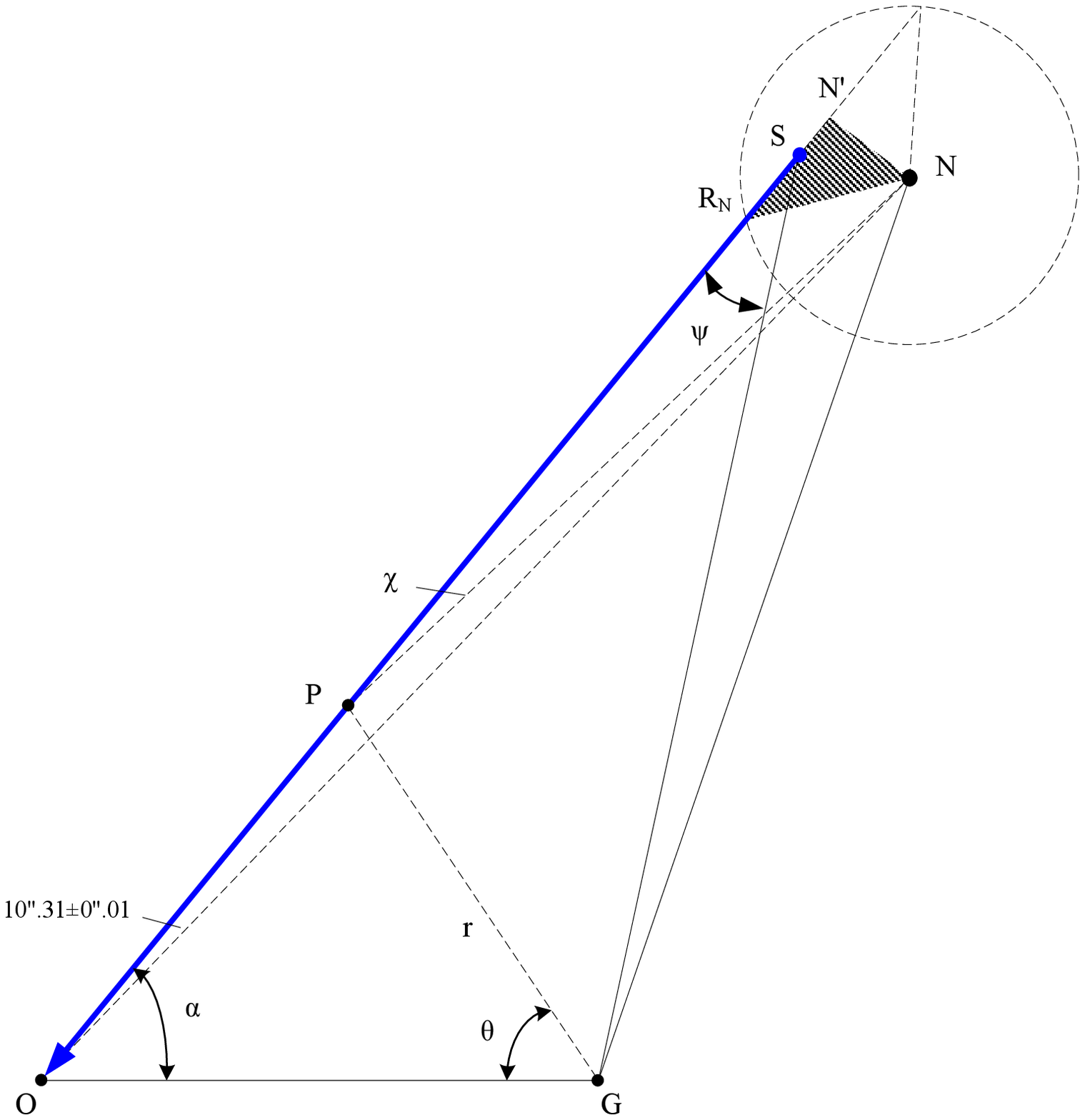}
\caption{\scriptsize The geometry of the travel path of waves from the merge position. The left panel shows the localization of the merge position (point $S$) in ECS, and the right panel shows the travel path of the waves (blue line) in PCS. The centers of the MW and NGC 4993 are denoted by points $G$ and $N$, respectively.}
\label{fig1-2}
\end{figure}

Here we present the geometry of the traveling path of the waves. We use ECS to describe the localization of the merge position (see the left panel of Figure \ref{fig1-2}), and use the polar coordinate system (PCS) to describe the path of the waves (see the right panel in Figure \ref{fig1-2}). In ECS, the impact parameter $b$ and the viewing angle $\alpha$ satisfy the formula as follows:
\begin{equation}
\label{parabalpha} b = SG \left(1 - \frac{SG^2}{4r_{G}^2}\right)^{1/2}, \ \ \ \cos \alpha = 1 - \frac{1}{2} \left(\frac{SG}{r_G}\right)^2.
\end{equation}
The distance $|SG|$ between points on the spherical surface can be given by
\begin{equation}\label{distwopoint}
|SG| = \left[(x_S - x_G)^2 + (y_S - y_G)^2 + (z_S - z_G)^2\right]^{1/2},
\end{equation}
where the coordinates $(x_S, y_S, z_S)$ and $(x_G, y_G, z_G)$ are shown as
\begin{eqnarray}
 x_S &=& r_G \cos \delta_S \cos \beta_S;\ \  x_G = r_G \cos \delta_G \cos \beta_G;\\
 y_S &=& r_G \cos \delta_S \sin \beta_S;\ \  y_G = r_G \cos \delta_G \sin \beta_G; \\
 z_S &=& r_G \sin \delta_S;\ \ \ \ \ \ \ \ \ \ z_G = r_G \sin \delta_G.
\end{eqnarray}
Here $r_G$ is assumed to be the radius of the celestial sphere.

Substituting $SG$ (\ref{distwopoint}) into Equation (\ref{parabalpha}), we get the formula below about the angle $\alpha$ between the line of sight and the line from Earth to the Galactic center,
\begin{equation}\label{201alpha}
\cos \alpha = \sin \delta_S \sin \delta_G + \cos \delta_S \cos \delta_G \cos \Delta \beta,
\end{equation}
with $\Delta \beta = |\beta_S-\beta_G|$. Then, by substituting the coordinates of points $S$ and $G$ into Equation (\ref{201alpha}), we get $\alpha \approx 61.28^o$. The impact parameter $b$ can thus be rewritten as
\begin{equation}\label{201IP}
 b^2 = r_G^2 \left[1 - \left(\sin \delta_S \sin \delta_G + \cos \delta_S \cos \delta_G \cos \Delta \beta\right)^2\right].
\end{equation}

In PCS, at the initial time of traveling, the wave is located at point $S$ $(r_S, \theta_S)$
\begin{eqnarray}
\label{201OGC11} r_S^2 &=& d^2 + r_G^2 - 2 r_G d \cos \alpha,\\
\label{201OGC22}d^2 &=& r_S^2 + r_G^2 - 2 r_G r_S \cos \theta_S,
\end{eqnarray}
with $r_S = GS$ and $d = OS$. In this way, we obtain the angles $\psi = 0.01^o$ and $\theta_S  = 118.71^o$ in $\triangle SOG$ at the initial time of wave's propagation. The path of the waves from the source $(r_S, \theta_S)$ to the final receiver $(r_G, 0^o)$ is illustrated in Figure \ref{fig1-2}. For any test point $P$ with the coordinate $(r, \theta)$, the angle $\alpha$ should satisfy the following formula:
\begin{equation}\label{psi2}
\cos \alpha = \frac{OP^2 +r_G^2 - r^2}{2 r_G \cdot OP}.
\end{equation}

Therefore, the dynamic distance from our Earth to any position $P$ during waves traveling can be obtained as,
\begin{equation}\label{pathcondition}
OP = \frac{1}{2 d} \left[\zeta \pm \sqrt{\zeta^2 + 4 d^2 \left(r^2 - r_G^2\right)} \right],
\end{equation}
where we adopt $\zeta = d^2 - r_S^2 + r_G^2$, and keep the sign ``+" in front of the square root. When the waves propagate along the path, it requires that at the initial moment of $r \rightarrow r_S$, the condition of $OP \rightarrow d$ must be satisfied, and at the terminal moment of $r \rightarrow r_G$, the condition of $OP \rightarrow 0$ also must be satisfied. The line $OP$ in Equation (\ref{pathcondition}) with sign ``+" is the path defined as the propagation of waves from the merge position to that of Earth.

In order to distinguish the potentials between the two galaxies, we use $r$ and $\chi$ to denote the radius of the MW and NGC 4993, respectively, in Equations (\ref{addmw}) and (\ref{addNGC}). If the waves travel along their path, the below condition should be upheld:
\begin{equation}\label{addppnn}
\chi(r, \theta) = \left[r^2 + GN^2 - 2 GN r \cos \left(\theta_S - \theta\right)\right]^{1/2},
\end{equation}
where, for one cosmic source of GW170817, we can assume $GN \approx d$ and $\theta_S \approx \angle NGO$.
\section{The angle offset}\label{app-2}
We then present the geometry of the observed angle offset. According to the projected triangle $\triangle PNN'$ in Figure \ref{fig1-2}, a key relationship about the path of the waves can be given as
\begin{equation}\label{addPN}
NP = \sqrt{N'N^2 + N'P^2},
\end{equation}
where $N'$ is the projected position of the galactic center (point $N$) along the line of sight. The projected offset distance $N'N$ can be used quantitatively to indicate the observed angle offset. When the angle offset disappears, i.e., $N'N \rightarrow 0$, the distance between the two galaxies approximates the distance from the merge position to Earth, i.e., $ON' \rightarrow ON$.

In Figure \ref{fig1-2}, the minimum of $SN'$ ($SN'=0$) comes from the fact that the source is located at the projected point $N'$. Inversely, the maximum of $SN'$ ($SN' = 13.0_{-4.6}^{+2.6}$~kpc) appears when the source is located at the front outermost edge (the bulk denoted by the dash line) of the galaxy along the line of sight. This also means that the source is located at the point $R_N$ where $R_N N = R_N$ is the half of the galaxy diameter. By using the NASA Extragalactic Database or the ESO-LV catalog \citep[see][]{Lauberts}, one can find that the diameter of NGC 4993 is about 26~kpc, which is longer than the bulge scale length ($\sim 1.5$~kpc) in the Hernquist profile. These magnitudes are consistent with each other. The reason for this is that the diameter represents the maximum range of possible source locations, and the bulge scale length represents the range of the main stellar mass producing a stellar potential (\ref{21stelpoten}). Therefore, we can obtain the minimum and the maximum $N'S$ shown by
\begin{eqnarray}
\label{addminnns}  N'S|_{min} &=& 0,\\
\label{addmaxnns}  N'S|_{max} &=& N'R_N = \sqrt{R_N^2 - N'N^2}.
\end{eqnarray}
The range of $N'S$ thus is determined by the distance $d$ from the source $S$ to the receiver $O$. The uncertainty in the diameter of NGC 4993, 26~kpc, is proportional to the uncertainty on the distance d = $40^{+8}_{-14}$~Mpc, so the uncertainty on the maximum of $N'S$ is proportional to this as well. Therefore, we can obtain the distance $N'S$ in the range of [0, $13.0_{-4.6}^{+2.6}$~kpc] where the maximum uncertainty comes from the luminosity distance.

\end{document}